\documentstyle[12pt]{article}
\newcommand{\be}{\begin{equation}}
\newcommand{\ee}{\end{equation}}
\begin{document}
\vspace*{0.5\baselineskip}
\begin{center}
{\large\bf Ionization of Rydberg atoms in a low frequency field: modelling by
maps of transition to chaotic behavior}\\[\baselineskip]
B. Kaulakys and G. Vilutis\\
{\small Institute of Theoretical Physics and Astronomy, A. Go\v stauto 12,
2600 Vilnius, Lithuania}\\[\baselineskip]
\parbox{5.5in}{{\bf Abstract.} We investigate a microwave ionization of highly
excited atom in a low frequency field and show that such a process may be
studied on the bases of map for the electron energy change during the period
of the electron motion between two subsequent passages at aphelion. Simple
approximate criterion results to the threshold field for transition to chaotic
behavior very close to the numerical results. We show that transition from
adiabatic to chaotic ionization mechanism takes place when the field frequency
to the electron's Kepler frequency ration approximately equals 0.1.}
\end{center}
\vskip 2\baselineskip

{\large 1. Introduction}
\vskip\baselineskip

Highly excited atom in a monochromatic field is one of the simplest real
strongly driven by an external driving field nonlinear system with stochastic
behavior. That is why recently a large amount of effort, both theoretical and
experimental, has been devoted to the studies of the Rydberg atoms in strong
microwave fields (for review see [1-3] and references therein). The
observation of excitation and ionization rates provides evidence for stochastic
behavior of weakly bound electron: the ionzation of Rydberg atoms exhibits a
threshold dependence on the field amplitude and, at least for the low relative
frequency of the field, appears as a diffusion like process. The classical
dynamics of the excited electron may be described by the map, called the
"Kepler
map", rather than by differential equations of motion [4,5], which greatly
facilitates the numerical and analytical investigation of stochasticity and
ionization process. The map, mostly represented for the number of absorbed
photons, is widely used for the analysis of chaotic processes for relatively
high frequencies of the field [1-5]. In the very low frequency region the
ionization is direct: the electron escapes over the potential barrier [5-6].
However, the threshold field for the direct low frequency (adiabatic)
ionization is considerably higher than that for the transition to chaotic
behavior
in the medium and high frequency field. On the other hand, the derivation of
the mapping equations of motion [4,5] is based on the classical perturbation
theory. Therefore, the question of the validity of these maps for description
of the ionization process in the low frequency relatively strong field arises.
In addition, the transition from adiabatic to chaotic ionization mechanism is
of great interest, especially as the classical ionization theory is proper
for the low frequency ionization.

It is the purpose of this work to investigate the fitness of the classical
maps [4,5] for the low-frequency ionization of Rydberg atoms and on this basis
to analyse a transition from adiabatic to chaotic motion in the strongly
driven by the slow external field system.

\vskip \baselineskip
{\large 2. Equations of motion and dynamics}
\vskip \baselineskip
The direct way of coupling the electromagnetic field to the electron
Hamiltonian is through the ${\bf A\cdot P}$ interaction, where ${\bf A}$ is
the vector potential of the field and ${\bf P}$ is the generalized momentum of
the electron. Thus the Hamiltonian of the hydrogen atom in a linearly
polarized field $F\cos(\omega t+\vartheta)$, where $F$ and $\omega$ are the
field strength amplitude and frequency, in atomic units has the form
\be
H={1\over 2}\left({\bf P}-{{\bf F}\over\omega}\sin(\omega t+\vartheta)
\right)^2-{1\over r}.
\ee
The electron energy change due to the interaction with the external field
follows from the Hamiltonian equations of motion [7]
\be
\dot E=-{\bf\dot r\cdot F}\cos(\omega t+\vartheta).
\ee
Measuring the time of the field action in the field periods one can introduce
the scale transformation [8] when the scaled field strength and energy are
$F_s=F/\omega^{4/3}$, and $E_s=E/\omega^{2/3}$. However, it is convenient to
introduce the positive scaled energy $\varepsilon=-2E_s$ and relative field
strength $F_0=Fn^4_0=F_s/\varepsilon_0^2$, with $n_0$ being the initial
principal quantum number.

Note, that eq. (2) is exact if ${\bf\dot r}$ is obtained from the equations of
motion including the influence of the electromagnetic field. Using the
parametric equations of motion we can calculate the change of the electron's
energy in the classical perturbation theory approximation.
We restrict our subsequent consideration to the one-dimentional model, which
corresponds to the states with low orbital quantum numbers $l\ll n$. The
integration
of eq. (2) for the motion between two subsequent passages at the aphelion
(where
$\dot x=0$ and there is no energy change) results to the map (see [4,5] for
details)
\be
\left\{
\begin{array}{rcl}
\varepsilon_{j+1}&=&\varepsilon_j-4\pi F_0\varepsilon_0^2
\varepsilon^{-1}_{j+1}{\bf J}'_{s_{j+1}}(s_{j+1})\sin\vartheta_j\\
\vartheta_{j+1}&=&\vartheta_j+
2\pi\varepsilon^{-3/2}_{j+1}+G(\varepsilon_{j+1},\vartheta_j)
\end{array}
\right.
\ee
where $s=\varepsilon^{-3/2}=\omega/(-2E)^{3/2}$ is the relative frequency of
the field, ${\bf J}'_s(z)$ is the derivative of the Anger function and function
$G$
may be obtained from the requirement of area-preserving [4]. In the low
frequency $s\ll 1$ limit ${\bf J}'_s(s)\simeq s/2$ and we have the map
\be
\left\{
\begin{array}{rcl}
\varepsilon_{j+1}&=&\varepsilon_j-2\pi F_0\left(\varepsilon_0^2/
\varepsilon^{5/2}_{j+1}\right)\sin\vartheta_j\\
\vartheta_{j+1}&=&\vartheta_j+
2\pi/\varepsilon^{3/2}_{j+1}+5\pi
F_0\left(\varepsilon_0^2/\varepsilon_{j+1}^{7/2}
\right)\cos\vartheta_j
\end{array}
\right..
\ee

The results of the numerical analysis of maps (3) and (4) in the low
frequency, $s\le
1$, limit are presented in Fig. 1 and 2.
\begin{table}
\vspace*{9in}
\end{table}
\begin{table}
\vspace*{10cm}
Fig. 2. The relative threshold field strengths for the ionization outset from
the numerical analysis of the maps (3) and (4) and according to the approximate
criterion (8)-(9)
\end{table}
We see that the threshold ionization
field approaches the static field ionization threshold $F_0^{st}\approx 0.13$
when $s_0\to 0$ ($\varepsilon\to\infty$). Thus, the maps (3) and (4) are valid
in the low frequency limit when the strength of the driving electric field is
of the order of the Coloumb field.

For the low frequencies, $2\pi s=2\pi/\varepsilon^{3/2}\ll 1$, the change of
the
angle $\vartheta$ after one step of iteration is small and we can transform the
system of equations (4) to the differential equation
\be
{d(\cos\vartheta)\over d\varepsilon}={\varepsilon\over\varepsilon_0^2F_0}+
{5\cos\vartheta\over 2\varepsilon}.
\ee
The analytical solution of eq. (5) with the initial conditions $\varepsilon=
\varepsilon_0$ when $\vartheta=\vartheta_0$ is
\be
\cos\vartheta=z^5\cos\vartheta_0-(2/F_0)z^4(1-z),~~~z=\sqrt{\varepsilon/
\varepsilon_0}
\ee

For the relatively low values of $F_0$ there is a motion in all the interval
$0\div 2\pi$ of the angle $\vartheta$. However for $F_0=2z^4/5$ the increase of
$\vartheta$ at $\vartheta\simeq\pi$ changes to the decrease and results to the
fast decrease of $\varepsilon$ and ionization process. The minimal value of
$F_0$ for such a motion corresponds to $\vartheta_0=0$ and may be defined as a
maximal value of $F_0$ resulting to the motion in the interval $0\div 2\pi$,
i. e., the maximum of the expression
\be
F_0=2z^4(1-z)/(1+z^5).
\ee
Such a maximum is at $z=z_0$, where $z_0$ is a solution of equation
$z^5+5z-4=0$,
i. e., $z_0=0.75193$ and results to the $F_0^0=0.1279$ which is only $1\%$
lower the adiabatic ionization value $F_0^{st}=2^{10}/(3\pi)^4\simeq 0.1298$.

The approximate criterion for transition to chaotic behavior is [5]
\be
K=\max|\delta\vartheta_{j+1}/\delta\vartheta_j-1|\ge 1
\ee
and according to eqs. (4) yields to the expression for the threshold field
strength
\be
F_0^c=\varepsilon^5/6\pi^2\varepsilon_0^2=z_c^{10}/6\pi^2s_0^2
\ee
where $z_c$ is the solution of eq. (7) with $F_0=F_0^c$. Note that
$z_c^{10}\simeq 1-10F_0^c+30(F_0^c)^2\mp\dots$ if $F_0^c\le 0.1$. For
$0.09\le s_0\le 0.5$ expression (9) gives the ionization threshold field
very close to the numerical results (see Fig. 2).
\vskip\baselineskip
{\large 3. Conclusion}
\vskip\baselineskip
{}From the above analysis we can conclude that the map at aphelion (3) is
suitable for the investigation of transition to chaotic behavior and ionization
of Rydberg atoms also in the low frequency field, even for adiabatic
ionization,
when the strength of the external field is comparable with the Coloumb field.
For such a purpose there is no need to use the map for the two halves of
intrinsic
period [9]. Moreover, the approximate criterion (8) for transition to chaotic
behavior yields to the threshold field strength very close to the numerical
results if we take into account the increase of the electron's energy by the
influence of the electromagnetic field. The transition from adiabatic to
chaotic
ionization occurs at the relative field frequency $s_0\simeq 0.1$.
\vskip\baselineskip
{\large Acknowledgement}
\vskip\baselineskip
The research described in this publication was made possible in part by Grant
No. LAA000 from the International Science Foundation.
\vskip\baselineskip
{\large Refereces}
\vskip\baselineskip
\begin{enumerate}
\item G. Casati, I. Guarneri and D. L. Shepelyanski, {\it IEEE J. Quant.
Electron.}, {\bf 24}, 1420 (1988).

\item R. V. Jensen, S. M. Suskind and M. M. Sanders, {\it Phys. Reports},
{\bf 201}, 1 (1991).

\item P. M. Koch in {\it Chaos and Quantum Chaos}, Lecture Notes in Physics
Vol. 411, ed. W. Dieter Heiss (Springer-Verlag, 1992), p. 167-224.

\item V. Gontis and B. Kaulakys, deposited in VINITI as No. 5087-V86, (1986)
and {\it Lit. fiz. sb. (Sov. Phys.-Collect.)}, {\bf 27}, 368 (111) (1987).

\item V. Gontis and B. Kaulakys, {\it J. Phys. B: At. Mol. Phys.}, {\bf 20},
5051
(1987).

\item R. V. Jensen, {\it Phys. Rev. A}, {\bf 30}, 386 (1989).

\item L. D. Landau and E. M. Lifshitz, {\it Classical Field Theory} (New York:
Pergamon, 1975).

\item B. Kaulakys {\it et al.}, {\it Phys. Lett. A} {\bf 159}, 261 (1991); {\it
Lith. Phys. J.}, {\bf 33}, 354 (290) (1993).

\item B. Kaulakys, {\it J. Phys. B: At. Mol. Opt. Phys.}, {\bf 24}, 571 (1991).
\end{enumerate}
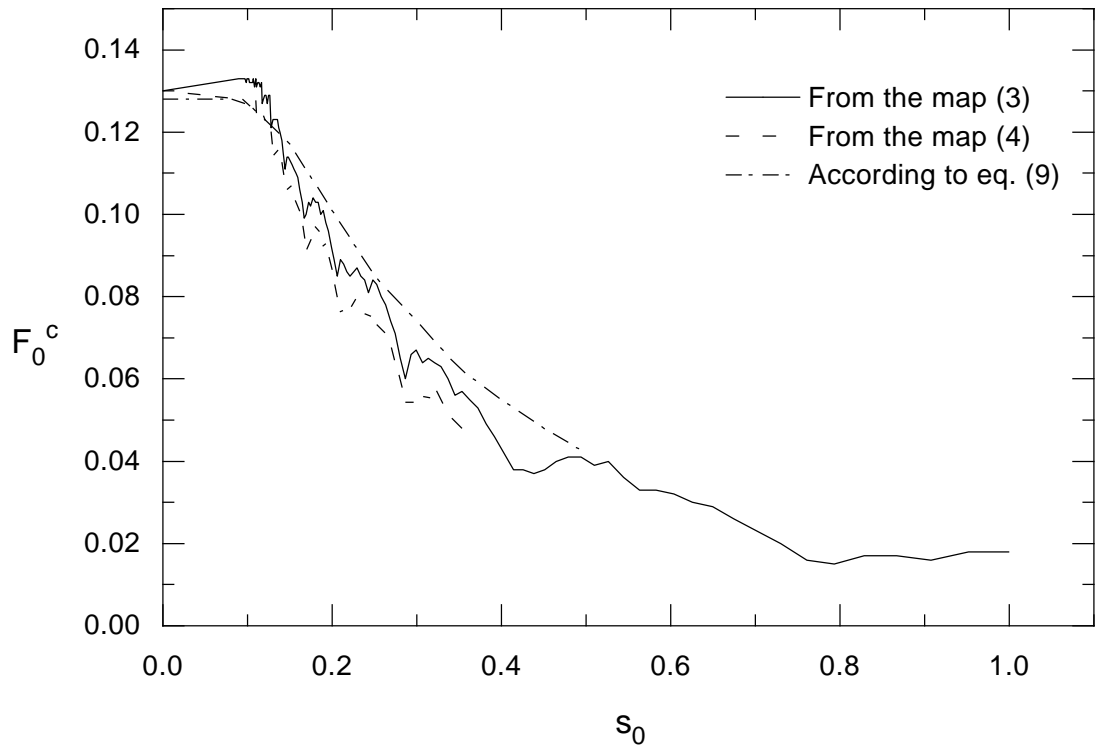
\end{document}